\documentclass{article}
\usepackage{times}
\usepackage{helvet}
\usepackage{courier}
\usepackage{graphicx}

\newcommand{\voteTotal}{N_{\mbox{\scriptsize vote}}}
\newcommand{\visitRate}{\nu} 
\newcommand{\frontRate}{\nu_{\mbox{\scriptsize f}}} 
\newcommand{\newRate}{\nu_{\mbox{\scriptsize u}}} 
\newcommand{\friendsRate}{\nu_{\mbox{\scriptsize friends}}} 
\newcommand{\fractionToPage}{f_{\mbox{\scriptsize page}}}
\newcommand{\frontPageGrowth}{k_{\mbox{\scriptsize f}}} 
\newcommand{\newPageGrowth}{k_{\mbox{\scriptsize u}}} 

\newcommand{\expect}[1]{\langle #1 \rangle}

\newcommand{\erfc}{\mbox{erfc}} 
\newcommand{\rms}{{\sc rms}} 

\newcommand{\hour}{\mbox{hr}} 
\newcommand{\minute}{\mbox{min}}

\title{Stochastic Models of User-Contributory Web Sites}

\author{Tad Hogg\\{\small Hewlett-Packard Laboratories} \and Kristina Lerman\\{\small USC Information Sciences Institute}}
\begin{document}
\maketitle

\begin{abstract}
We describe a general stochastic processes-based approach to modeling
user-contributory web sites, where users create, rate and share
content. These models describe aggregate measures of activity and
how they arise from simple models of individual users. This approach
provides a tractable method to understand user activity on the web
site and how this activity depends on web site design choices,
especially the choice of what information about other users' behaviors
is shown to each user. We illustrate this modeling approach in the
context of user-created content on the news rating site Digg.
\end{abstract}

\section{Introduction}

The Web is becoming more complex and dynamic as sites allow users to
contribute and personalize content. Such sites include Digg, Flickr
and YouTube where users share and rate news stories, photos and
videos, respectively. Additional examples of such web sites include
Wikipedia and Bugzilla, enabling anyone to contribute to
encyclopedia articles or help develop open source software. These
social web sites also often allow users to form explicit links with
other users whose contributions they find interesting and highlight
the activity of a user's designated friends~\cite{lerman08} to help
users find relevant content.

Web sites often provide users with aggregate summaries of recent
activity. For example, both Digg and Flickr have a front page that
features `hot' (popular or interesting) content. News organizations,
such as The New York Times, allow users to subscribe to or embed RSS
feeds of their most popular (e.g., emailed) stories in the users'
own pages. Feedback between individual and collective actions can
lead to nonlinear amplification of even small signals. For example,
the `Digg effect' refers to the phenomenon where a `hot' story on
the social news aggregator Digg brings down servers hosting the
story that are not equipped to deal with heavy traffic that a
popular story on Digg generates.

Aggregate activity of many users determines the structure and
usefulness of user-participatory web sites. Understanding this
emergent behavior will enable, for example, predicting which newly
contributed content will likely become popular, identifying
productive ways to change how information is displayed to users, or how to
change user incentives so as to improve the content.

The behavior of an individual user on a user-contributory web site is governed by a myriad of
social, economic, emotional and cognitive factors, and often subject
to unpredictable environmental influences, such as the weather or
the economy. Nevertheless, the combined activities of many
users often produce remarkably robust aggregate
behaviors~\cite{wilkinson08,wu07}.

In this paper, we present a stochastic processes-based framework for
relating aggregate behavior of web users to simple descriptions of
their typical individual behavior. The models can be written
directly from the individual behavior descriptions, and quantified
with empirical observations of a representative sample of users.

The methodology we describe applies to behaviors that can be modeled
as Markov processes, i.e., where the relevant changes depend only on
the current state of the system, not the detailed history of how it
arrived at that state. In principle such models can always be
applied by extending the complexity of the ``state'' describing the
system. However, such complexity can lead to models requiring
estimates for an impractically large number of parameters
characterizing how the state changes. Instead, the Markov modeling
assumption is useful primarily in connection with systems requiring
only a few variables to define their current state.

At first glance an assumption of Markov processes and simple states
may appear overly restrictive for describing human behavior.
However, many online activities provide only a fairly limited set of
actions for users and present information based on little or no
historical context of particular individuals. In these cases, a few
state variables can capture the main context involved in user
actions. Furthermore, we discuss simplifying approximations to the
models that readily enable identifying how key system behaviors
relate to user actions. These simplifications come at a cost: while
the resulting models correctly describe the typical aggregate
behaviors, they say little about their extreme cases, e.g., where
web site use is suddenly and briefly much larger than average. Even
with this limitation, however, simplified models are often preferred
over full models, which frequently require multiple simulation
trials, which are computationally expensive and whose typical
behaviors can be challenging to identify~\cite{lerman01}.

The paper is organized as follows. Section~\ref{sec:stochastic}
reviews the stochastic modeling framework. In Section~\ref{sec:digg}
we then illustrate the framework for the social news aggregator
Digg~\cite{lerman07}, which allows users to submit and rate news
stories by voting for the stories they like. Digg promotes highly
rated stories to the front page, in essence allowing it to emerge
from the decisions made by its users. We describe in detail the
modeling steps: specifying an individual's behavior on a site,
estimating model parameters and solving for aggregate behaviors.  We
show stochastic models can correctly explain several features of
this collective user behavior. Because other user-contributory web
sites have features and aggregate behaviors~\cite{wilkinson08,wu07}
similar to Digg, the stochastic modeling approach could be useful in
describing a variety of sites in addition to Digg. This paper also
provides a brief tutorial of useful guidelines for applying the
stochastic modeling framework to the behavior of user-participatory
sites.

\section{Stochastic Models}\label{sec:stochastic}

Rather than account for the inherent variability of individuals,
stochastic models focus on the behavior of {\em average} quantities
representing aggregate properties of the system. In the context of a
participatory web site, such quantities include average rate at
which users contribute new content and rate existing content. Such
macroscopic descriptions often have a simple form and are
analytically tractable. Stochastic models do not reproduce the
results of a single observation --- rather, they describe
\emph{typical} behavior. These models are analogous to the approach
used in statistical physics, demographics and macroeconomics where
the focus is on relations among aggregate quantities, such as volume
and pressure of a gas, population of a country and immigration, or
interest rates and employment.

We represent each user as a stochastic process with a small number
of states. This abstraction captures much of the relevant individual
user complexity by casting their decisions as inducing probabilistic
transitions between states. This modeling framework applies to
stochastic processes of varying complexity. In this paper, we focus
on simple processes that obey the Markov property, namely, a user
whose future state depends only on her present state and the input
she receives. A Markov process can be succinctly captured by a
diagram showing the possible states of the user and conditions for
transition between those states.

With the representation of users based on a small set of relevant
states, the same set of states for all users, and transitions
depending only on the state and not the individual user, the system
as a whole is described simply by the \emph{number} of users in each
state at a given time. That is, the system configuration is defined
by the occupation vector: $ {\vec n} = (n_1, n_2,\ldots) $ where
$n_k$ is the number of users in state $k$.

The occupation vector changes as people use the web site, e.g., to
view, post and rate content. In principle, one could follow the
history of the system, giving a sequence of occupation vectors.
However, to investigate typical behavior we consider a collection of
histories of similar content (as determined through a few
characteristic properties). This grouping allows the model to
generalize from simply describing what has already been observed to
predicting behavior of similar situations that may arise in the
future.

The next step in developing the stochastic model summarizes the
variation within the collection of histories with a probabilistic
description. That is, we characterize the possible occupation
vectors by the probability, $P({\vec n},t)$, the system is in
configuration ${\vec n}$ at time $t$.
The evolution of $P({\vec n},t)$ is given by the Stochastic Master Equation~\cite{vankampen92}.

Solving the Master Equation analytically is almost always
intractable. Monte Carlo simulations can determine the model's
predictions, and are often feasible for evaluating relatively small
group behavior~\cite{steglich07,robins07}. However, simulations are
computationally challenging for large groups, such as the thousands
to millions of users of contributory web sites. Moreover, the
requirement to repeat the simulation many times to identify typical
behavior makes it difficult to identify the key features of
information available to users and their choices leading to the
observed aggregate behavior of the web site. An alternative is a
simple, but approximate, method working with the average occupation
number, whose evolution is given by the Rate Equation
\begin{equation}
\label{eqn-rate-3} \frac {d \expect{n_k}} {d t} = \sum_{j}
w_{jk}(\expect{\vec n}) \expect{n_{j}} - \expect{ n_k}
\sum_{j}w_{kj}(\expect{ \vec n})
\end{equation}
where $\expect{ n_k}$ denotes the average number of users in state
$k$ at time $t$, i.e., $\sum_{\vec n}n_k P({\vec n}, t)$, and
$w_{jk}(\expect{\vec n})$ is an approximate expression for the
average transition rate, $\expect{w_{jk}(\vec n)}$, from
configuration $j$ to configuration $k$, with $w_{jk}(\vec n)$ the
transition rate when the occupation vector is $\vec n$. The
transition rates can also depend explicitly on time. In the Rate
Equation, occupation number $n_k$ increases due to users'
transitions from other states to state $k$, and decreases due to
transitions from the state $k$ to other states. Each state
corresponds to a dynamic variable in the mathematical model --- the
average number of users in that state --- and it is coupled to other
variables via transitions between states.

Using the average of the occupation vector in the transition rates,
i.e., $w_{jk}(\expect{\vec n})$, rather than the average of the
transition rates for the possible occupation vectors, i.e.,
$\expect{w_{jk}(\vec n)}$, is a common simplifying technique for
stochastic models. A sufficient condition for the accuracy of this
approximation is that variations around the average are relatively
small, so the average is a fair description of the typical behavior.
In stochastic models with many components, variations are often
small due to many independent interactions among the components.
More elaborate versions of the stochastic approach give improved
approximations when variations are not small, particularly due to
correlated interactions~\cite{opper01}. Even in these cases,
however, the averaged rate equations usually provide useful
qualitative relations between user behaviors and aggregate
properties of the system. User behavior on the web often involves
distributions with long tails, whose typical behaviors differ
significantly from the average~\cite{wilkinson08}. In this case we
have no guarantee that the averaged approximation is adequate.
Instead we must test its accuracy for particular aggregate behaviors
by comparing model predictions with observations of actual behavior,
as we report below.

In summary, this stochastic modeling approach to typical aggregate
behavior requires specifying the aggregate states of interest and
how individual user actions cause transitions among these states.
The modeling approach is best suited to cases where the users'
decisions are mainly determined by a few characteristics of the user
and the information they have about the system. These system states
and transitions give the rate equations. Solutions to these
equations then give estimates of how aggregate behavior varies in
time and depends on the characteristics of the users involved.

The aggregate behavior described by the Rate Equations is universal,
i.e., the same formalism describes a variety of systems governed by
the same abstract principles. This approach successfully models
several distributed robot
systems~\cite{lerman01,martinoli04,galstyan05}. Stochastic models
also describe group behavior in social science, with parameters
estimated from social surveys~\cite{robins07}, e.g., the formation
of network connections among teen peer groups~\cite{steglich07}.

At the heart of this argument is the concept of separation of
scales, which holds that the details of microscopic (user-level)
interactions are only relevant for computing the values of
parameters of the macroscopic model. This principle applies broadly
to naturally evolved systems, as found in biology and economics, and
designed technological artifacts~\cite{courtois85,simon96}. From the
perspective of large-scale group behaviors, this decomposition often
arises from processing, sensory and communication limitations of the
individuals and their limited set of actions. In effect, these
limits mean users can only pay attention to a relatively small
number of variables~\cite{hogg87PhysRep}.

\section{Example: Stochastic Model of Digg}\label{sec:digg}

As an example of stochastic modeling, we examine aggregate behavior
on Digg, a social news aggregator whose users submit and rate
stories. When a user submits a story, it goes to the upcoming
stories list. There are a few new submissions every minute and they
are displayed in reverse chronological order of their submission
time, 15 stories to a page. A user votes on a story by ``digging''
it. A newly submitted story is visible on the upcoming stories pages
for 24 hours after the submission. If the story accumulates enough
votes within this 24-hour period, it is promoted to the front page,
and becomes visible there. Otherwise, the story is removed after 24
hours. Although the exact promotion mechanism is kept secret and
changes occasionally, it appears to take into account the number of
votes the story receives and how rapidly. Digg's popularity is
fueled in large part by the emergent front page.

Digg allows users to track friends' activities (stories they
recently submitted or voted for). The friend relationship is
asymmetric. When user $A$ lists user $B$ as a \emph{friend}, $A$ can
watch the activities of $B$ but not vice versa. We call $A$ the
\emph{fan} of $B$. A submitted story is visible in the upcoming
stories list, as well as to submitter's fans through the Friends
interface. With each vote, a story becomes visible to the voter's
fans through the ``dugg upcoming'' part of the friends interface,
which shows the newly submitted stories that user's friends voted
for.

While in the upcoming stories list, a story accrues votes slowly.
Once it is promoted to the front page, it accumulates votes at a
much faster rate. As the story ages, accumulation of new votes slows
down~\cite{wu07}, and after a few days the story's number of votes
saturates.

We collected data by scraping Digg over a period of several days in
May 2006. We determined the number of diggs for the stories as a
function of the time since each story's submission. We collected at
least 4 such observations for each of 2152 stories, submitted by
1212 distinct users. Of these stories, 510, by 239 distinct users,
were promoted to the front page. To focus on promoted stories, we
sampled more extensively from the front pages, so our data set has a
larger fraction of promoted stories than Digg as a whole.
We also determined the number of fans of a subset of users,
including those who submitted the stories we followed.

In this section we illustrate the stochastic approach with a model
describing how the number of votes received by stories changes in
time and depends on parameters characterizing individual user
behavior, on average. Our goal is to produce a model that explains
--- and predicts --- the voting patterns on Digg and
how these aggregate behaviors relate to the ways Digg enables users
to discover new content.

\begin{figure}[t]
  \begin{center}
  \includegraphics[height=1.9in]{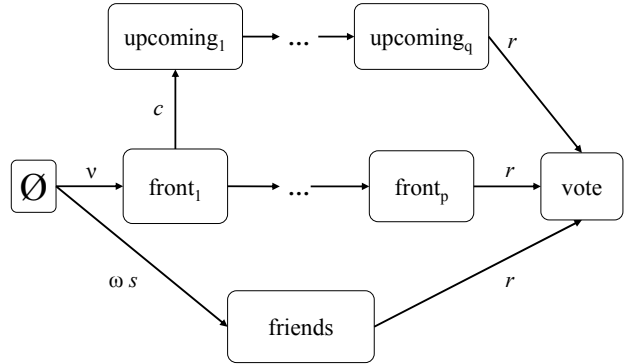}
  \end{center}
\caption{State diagram of user behavior for a single story. A user
starts in the $\emptyset$ state at the left, may find the story
through one of the three interfaces and may then vote on it. At a
given time, the story is located on a particular page of either the
upcoming or front page lists, not both. This diagram shows votes for
a story on either page $p$ of the front pages or page $q$ of the
upcoming pages. Only fans of previous voters can see the story
through the friends interface. Users in the friends, front or
upcoming states may choose to leave Digg, thereby returning to the
$\emptyset$ state (with those transitions not shown in the figure).
Users reaching the ``vote'' state remain there indefinitely and can
not vote on the story again. Parameters next to the arrows
characterize state transitions.} \label{fig:user-fsa}
\end{figure}

\subsection{Behavioral Model}

A user visiting Digg can choose to browse its \emph{front} pages to
see the recently promoted stories, \emph{upcoming} stories pages for
the recently submitted stories, or use the \emph{friends} interface
to see the stories her friends have recently submitted or voted for.
She can select a story to read and, if she considers it interesting,
\emph{vote} for it. Fig.~\ref{fig:user-fsa} shows the state diagram
for user's behavior. The user's environment, the stories she is
seeing, changes in time due to the actions of all the users.

We characterize the changing state of a story by three values: the
number of votes, $\voteTotal(t)$, the story has received by time $t$
after it was submitted to Digg, the list the story is in at time $t$
(upcoming or front pages) and its location within that list, which
we denote by $q$ and $p$ for upcoming and front page lists,
respectively.

With Fig.~\ref{fig:user-fsa} as a modeling blueprint, we relate the
users' choices to the changes in the state of a single story. In
terms of the general rate equation (Eq.~\ref{eqn-rate-3}), the
occupancy vector ${\vec n}$ describing the aggregate user behavior
at a given time has the following components: the number of users
who see a story via one of the front pages, one of the upcoming
pages, through the friends pages, and number of users who vote for a
story, $\voteTotal$. Since we are interested in the number of users
who reach the vote state, we do not need a separate equation for
each state in Fig.~\ref{fig:user-fsa}: at a given time, a particular
story has a unique location on the upcoming or front page lists.
Thus, for simplicity, we can group the separate states for each list
in Fig.~\ref{fig:user-fsa}, and consider just the combined
transition for a user to reach the page containing the story at the
time she visits Digg. These combined transition rates depend on the
location of the story in the list, i.e., the value of $q$ or $p$ for
the story. With this grouping of user states, the rate equation for
$\voteTotal(t)$ is:
\begin{equation}\label{eq:diggs}
    \frac{d \voteTotal(t)}{d t} =r ( \frontRate(t) + \newRate(t) + \friendsRate(t) )
\end{equation}
\noindent where $r$ measures how interesting the story is, i.e., the
probability a user seeing the story will vote on it, and
$\frontRate$, $\newRate$ and $\friendsRate$ are the rates at which
users find the story via one of the front or upcoming pages, and
through the friends interface, respectively.

In this model, the transition rates appearing in the rate equation
depend on the time $t$ but not on the occupation vector.
Nevertheless, the model could be generalized to include such a
dependence if, for example, a user currently viewing an interesting
story not only votes on it but explicitly encourages people they
know to view the story as well.

\subsection{Story Visibility}

Before we can solve Eq.~\ref{eq:diggs}, we must model the rates at
which users find the story through the various Digg interfaces.
These rates depend on the story's location in the list. The
parameters of these models depend on user behaviors that are not
readily measureable. Instead, we estimate them using data collected
from Digg, as described below.

\paragraph{Visibility by position in list}
A story's visibility on the front page or upcoming stories lists
decreases as recently added stories push it further down the list.
The stories are shown in groups: the first page of each list
displays the 15 most recent stories, page 2 the next 15 stories, and
so on.

We lack data on how many Digg visitors proceed to page 2, 3 and so
on in each list. However, when presented with lists over multiple
pages on a web site, successively smaller fractions of users visit
later pages in the list. One model of users following links through
a web site considers users estimating the value of continuing at the
site, and leaving when that value becomes
negative~\cite{huberman98}. This model leads to an inverse Gaussian
distribution of the number of pages $m$ a user visits before leaving
the web site,
\begin{equation}\label{eq:stopping distribution}
e^{-\frac{\lambda  (m-\mu )^2}{2 m \mu ^2}} \sqrt{\frac{\lambda
   }{2 \pi m^3}}
\end{equation}
with mean $\mu$ and variance $\mu^3/\lambda$. This distribution
matches empirical observations in several web
settings~\cite{huberman98}. When the variance is small, for
intermediate values of $m$ this distribution approximately follows a
power law, with the fraction of users leaving after viewing $m$
pages decreasing as $m^{-3/2}$.

To model the visibility of a story on the $m^{th}$ front or upcoming
page, the relevant distribution is the fraction of users who visit
\emph{at least} $m$ pages, i.e., the upper cumulative distribution
of Eq.~\ref{eq:stopping distribution}. For $m>1$, this fraction is
\begin{equation}
\fractionToPage(m) = \frac{1}{2}\left( F_m(-\mu) - e^{2\lambda/\mu}
F_m(\mu) \right)
\end{equation}
where $F_m(x)=\erfc(\alpha_m (m-1+x)/\mu)$, $\erfc$ is the
complementary error function, and $\alpha_m =
\sqrt{\lambda/(2(m-1))}$. For $m=1$, $\fractionToPage(1)=1$.

The visibility of stories decreases in two distinct ways when a new
story arrives. First, a story moves down the list on its current
page. Second, a story at the $15^{th}$ position moves to the top of
the next page. For simplicity, we model these processes as
decreasing visibility, i.e., the value of $\fractionToPage(m)$,
through $m$ taking on fractional values within a page, i.e., $m=1.5$
denotes the position of a story half way down the list on the first
page. This model is likely to somewhat overestimate the loss of
visibility for stories among the first few  of the 15 items on a
given page since the top several stories are visible without
requiring the user to scroll down the page.

\begin{figure}[t]
\begin{center}
  \begin{tabular}{cc}
  \includegraphics[height=1.8in]{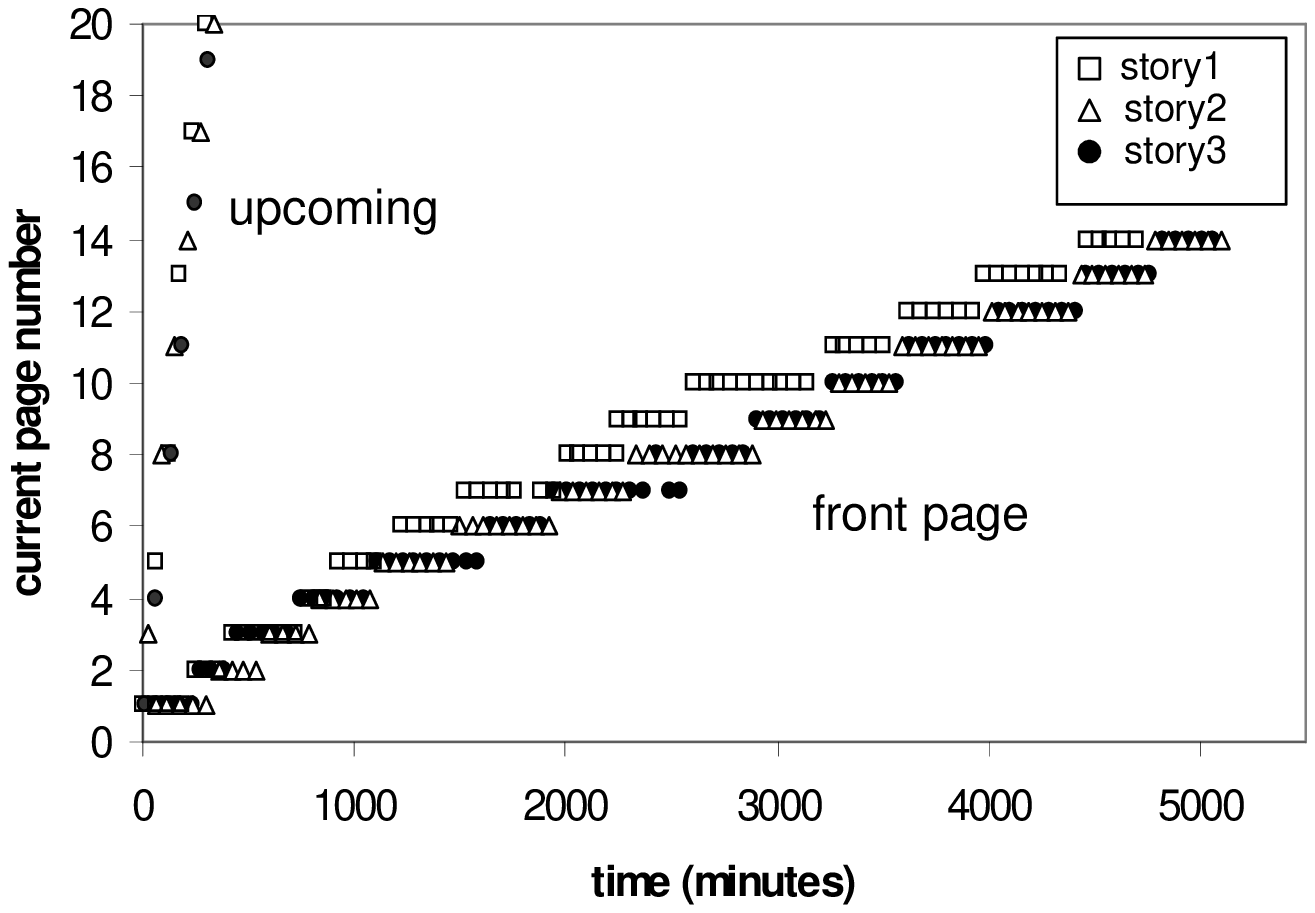} &
  \includegraphics[height=1.8in]{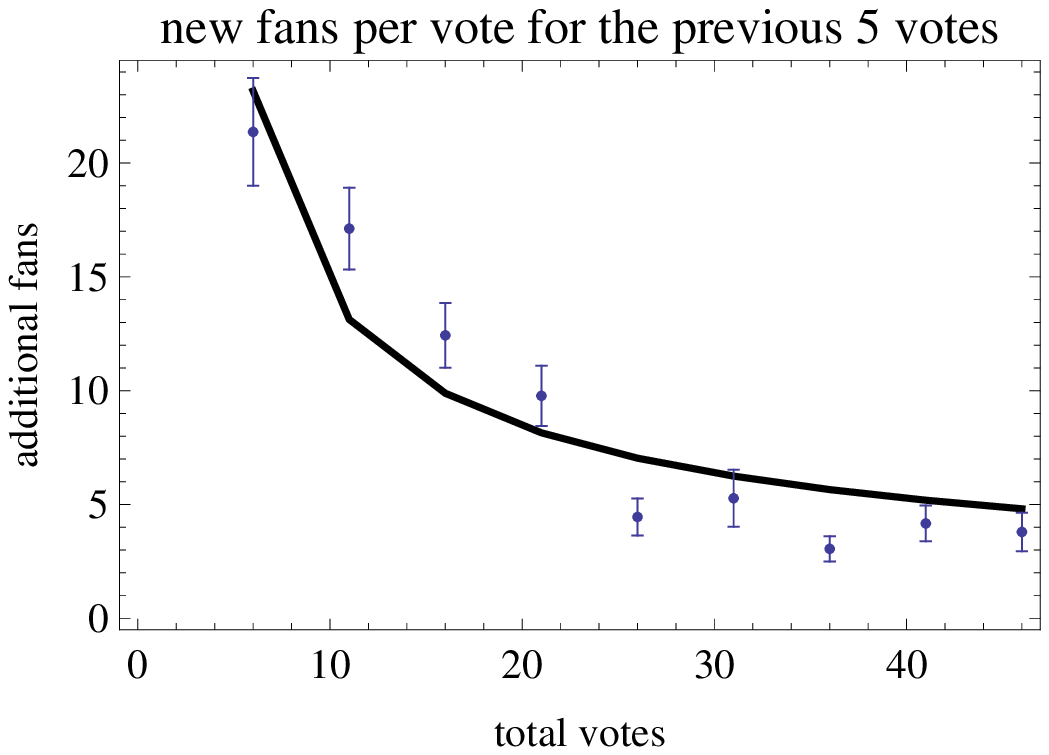} \\
  (a) & (b)
  \end{tabular}
\end{center}
\caption{(a) Current page number on the upcoming and front pages
vs.~time for three different stories. Time is measured from when the
story first appeared on each page, i.e., time it was submitted or
promoted, for the upcoming and front page points, respectively. (b)
Increase in the number of distinct users who can see the story
through the friends interface with each group of five new votes for
the first 46 users to vote on a story. The points are mean values
for 195 stories, including those shown in (a), and the curve is
based on Eq.~\ref{eq:Sv}. The error bars indicate the standard error
of the estimated means.} \label{fig:params}
\end{figure}

\paragraph{List position of a story}
Fig.~\ref{fig:params}(a) shows how the page number of a story on the
two lists changes in time for three randomly chosen stories from our
data set. The behavior is close to linear, so we take a story's page
number on the upcoming page $q$ and the front page $p$ at time $t$
to be\footnote{$\Theta(x)$ is a step function: $1$ when $x \ge 0$
and $0$ when $x<0$.}:
\begin{eqnarray}
  p(t) &=& (\frontPageGrowth (t-T_h)+1)\Theta(t-T_h) \\
  q(t) &=& \newPageGrowth t + 1
\end{eqnarray}
where $T_h$ is the time the story is promoted to the front page
(before which $p(t)=0$) and the slopes are given in
Table~\ref{parameters}.
Since each page holds 15 stories, these rates are $1/15^{th}$ the
submission and promotion rates, respectively.

\paragraph{Front page and upcoming stories lists}
Digg prominently shows the stories on the front page. The upcoming
stories list is less popular than the front page. We model this fact
by assuming a fraction $c<1$ of Digg visitors proceed to the
upcoming stories pages.

\paragraph{Promotion to the front page}
We use a simple threshold to model how a story is promoted to the
front page. Initially the story is visible on the upcoming stories
pages. If and when the number of votes a story receives exceeds a
promotion threshold $h$, the story moves to the front page. This
threshold model approximates Digg's promotion algorithm as of May
2006, since in our data set we did not see any front page stories
with fewer than 44 votes, nor did we see any upcoming stories with
more than 42 votes. We take $h=40$ as an approximation to the
promotion algorithm.

\paragraph{Friends interface}
The friends interface allows the user to see the stories her friends
have (i) submitted, (ii) voted for, and (iii) commented on in the
preceding 48 hours. Although users can take advantage of all these
features, we only consider the first two. These uses of the friends
interface are similar to the functionality offered by other social
media sites: e.g., Flickr allows users to see the latest images his
friends uploaded, as well as the images a friend liked.

The fans of the story's submitter can find the story via the friends
interface. As additional people vote on the story, their fans can
also see the story. We model this with $s(t)$, the number of fans of
voters on the story by time $t$ who have not yet seen the story.
Although the number of fans is highly variable, the average number
of additional fans from an extra vote when the story has
$\voteTotal$ votes is approximately
\begin{equation}\label{eq:Sv}
\Delta s = a \voteTotal^{-b}
\end{equation}
where $a=51$ and $b=0.62$, as illustrated in
Fig.~\ref{fig:params}(b), showing the fit to the \emph{increment} in
average number of fans per vote over groups of 5 votes as given in
the data. Thus early voters on a story tend to have more new fans
(i.e., fans who are not also fans of earlier voters) than later
voters.

The model can incorporate any distribution for the times fans visit
Digg. We suppose these users visit Digg daily, and since they are
likely to be geographically distributed across all time zones, the
rate fans discover the story is distributed throughout the day. A
simple model of this behavior takes fans arriving at the friends
page independently at a rate $\omega$. As fans read the story, the
number of potential voters gets smaller, i.e., $s$ decreases at a
rate $\omega s$, corresponding to the rate fans find the story
through the friends interface, $\friendsRate$. We neglect additional
reduction in $s$ from fans finding the story without using the
friends interface.

Combining the growth in the number of available fans and its
decrease as fans return to Digg gives
\begin{equation}
\frac{d s}{d t} = -\omega s + a \voteTotal^{-b} \frac{d
\voteTotal}{d t}
\end{equation}
with initial value $s(0)$ equal to the number of fans of the story's
submitter, $S$.
This model of the friends interface treats the pool of fans
uniformly. That is we assume no difference in behavior, on average,
for fans of the story's submitter vs.~fans of other voters.

\paragraph{Summary}
In summary, the rates in Eq.~\ref{eq:diggs} are:
\begin{eqnarray*}
  \frontRate &=&  \visitRate \fractionToPage(p(t)) \, \Theta(\voteTotal(t)-h) \\
  \newRate &=& c \, \visitRate \fractionToPage(q(t)) \, \Theta(h-\voteTotal(t)) \Theta(24\hour-t)\\
  \friendsRate &=& \omega s(t)
\end{eqnarray*}
\noindent where $t$ is time since the story's submission and
$\visitRate$ is the rate users visit Digg. The first step function
in $\frontRate$ and $\newRate$ indicates that when a story has fewer
votes than required for promotion, it is visible in the upcoming
stories pages; and when $\voteTotal(t)>h$, the story is visible on
the front page. The second step function in $\newRate$ accounts for
a story staying in the upcoming list for at most $24$ hours. We
solve Eq.~\ref{eq:diggs} subject to initial condition
$\voteTotal(0)=1$, because a newly submitted story starts with a
single vote, from the submitter.

\subsection{Model Parameters and Behavior}

\begin{table}[t]
\begin{center}
\begin{tabular}{l|l}
\hline parameter & value \\
\hline rate general users come to Digg & $\nu=10\,\mbox{users}/\minute$ \\
fraction viewing upcoming pages  & $c=0.3$ \\
rate a voters' fans come to Digg & $\omega=0.002/\minute$ \\
page view distribution  & $\mu=0.6$, $\lambda=0.6$ \\
fans per new vote & $a=51$, $b=0.62$ \\
vote promotion threshold    & $h=40$ \\
upcoming stories list location  & $\newPageGrowth = 0.06\,\mbox{pages}/\minute$ \\ 
front page list location  & $\frontPageGrowth = 0.003\,\mbox{pages}/\minute$ \\ 
\hline \multicolumn{2}{c}{story specific parameters} \\
interestingness    & $r$ \\
number of submitter's fans  & $S$ \\
\end{tabular}
\end{center}
\caption{Model parameters.}\label{parameters}
\end{table}

The solutions of Eq.~\ref{eq:diggs} show how the number of votes
received by a story changes in time. The solutions depend on the
model parameters, of which only two parameters
--- the story's interestingness $r$ and number of fans the submitter
has $S$ --- change from one story to another. Therefore, we fix
values of the remaining parameters as given in
Table~\ref{parameters}.

As described above, we estimate some of these parameters (such as
the growth in list location, promotion threshold and fans per new
vote) directly from the data. The remaining parameters are not
directly given by our data set (e.g., how often users view the
upcoming pages) and instead we estimate them based on the model
predictions. The small number of stories in our data set, as well as
the approximations made in the model, do not give strong constraints
on these parameters. We selected one set of values giving a
reasonable match to our observations. For example, the rate fans
visit Digg and view stories via the friend's interface, given by
$\omega$ in Table~\ref{parameters}, has 90\% of the fans of a new
voter returning to Digg within the next 19 hours.
As another example of interpreting these parameter values, for the
page visit distribution the values of $\mu$ and $\lambda$ in
Table~\ref{parameters} correspond to about $1/6$ of the users
viewing more than just the first page.
These parameters could in principle be measured independently from
aggregate behavior with more detailed information on user behavior.
Measuring these values for users of Digg, or other similar web
sites, could improve the choice of model parameters.

\begin{figure}[t]
\begin{center}
\includegraphics[width=3in]{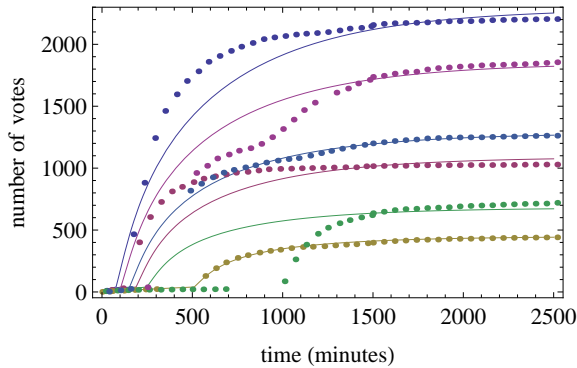}
\end{center}
\caption{Evolution of the number of votes received by six stories
compared with model solution. } \label{fig:predictions}
\end{figure}

\begin{table}[t]
\begin{center}
\begin{tabular}{ccc}
S & r & final votes \\
\hline 5 & 0.51 & 2229 \\
5 & 0.44 & 1921 \\
40 & 0.32 & 1297 \\
40 & 0.28 & 1039 \\
160 & 0.19 & 740 \\
100 & 0.13 & 458 \\
\end{tabular}
\end{center}
\caption{Parameters for the stories, listed in decreasing order of
total votes received by the story and hence corresponding to the
curves in Fig.~\ref{fig:predictions} from top to
bottom.}\label{story parameters}
\end{table}

As specific examples, we consider stories promoted to the front
page, though the model also describes the stories that never reach
the front page. Fig.~\ref{fig:predictions} shows the behavior of six
stories. For each story, $S$ is the number of fans of the story's
submitter, available from our data, and $r$ is estimated to minimize
the root-mean-square (\rms) difference between the observed votes
and the model predictions. Table~\ref{story parameters} lists these
values. The more interesting stories (with higher $r$ values) are
promoted to the front page (inflection point in the curve) faster
and receive more votes than less interesting stories. Overall there
is qualitative agreement between the data and the model, indicating
that the features of the Digg user interface we considered can
explain the patterns of collective voting. This highlights a benefit
of the stochastic approach: identifying simple models of user
behavior that are sufficient to produce the aggregate properties of
interest.

The only significant difference between the data and the model is
visible in the lower two lines of Fig.~\ref{fig:predictions}. In the
data, a story posted by the user with $S=100$ is promoted before the
story posted by the user with $S=160$, but saturates at smaller
value of votes than the latter story. In the model, the story with
larger $r$ is promoted first and gets more votes.

\begin{figure}[t]
\begin{center}
  \begin{tabular}{cc}
  \includegraphics[height=1.8in]{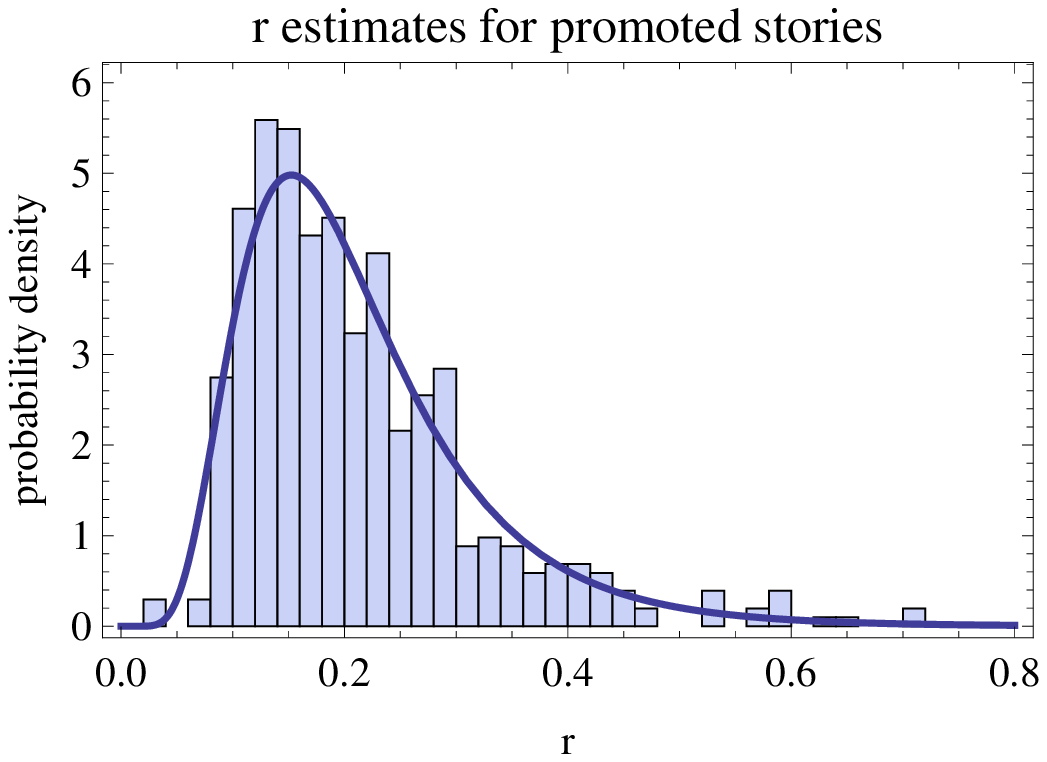} &
  \includegraphics[height=1.8in]{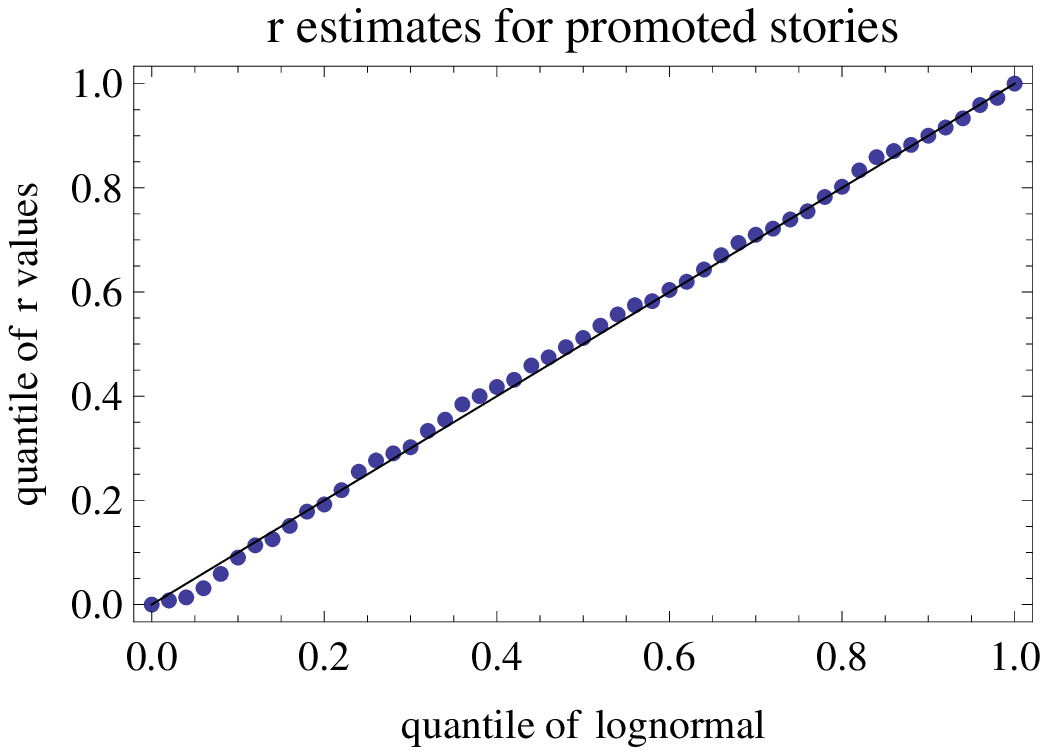} \\
  (a) & (b)
  \end{tabular}
\end{center}
\caption{(a) Histogram of estimated $r$ values for the promoted
stories in our data set compared with the best fit lognormal
distribution. (b) Quantile-quantile plot comparing observed
distribution of $r$ values with the lognormal distribution fit.}
\label{fig:r-distribution}
\end{figure}

The estimated $r$ values for the stories in our data set show the
510 promoted stories have a wide range of interestingness to users.
As shown in Fig.~\ref{fig:r-distribution}, these $r$ values fit well
to a lognormal distribution with maximum likelihood estimates of the
mean and standard deviation of $\log(r)$ equal to $-1.67\pm0.04$ and
$0.47\pm0.03$, respectively, with the ranges giving the $95\%$
confidence intervals. A randomization test based on the
Kolmogorov-Smirnov statistic and accounting for the fact that the
distribution parameters are determined from the
data~\cite{clauset07} shows the $r$ values are consistent with this
distribution ($p$-value $0.35$). While broad distributions occur in
several web sites~\cite{wilkinson08}, our model allows factoring out
the effect of visibility due to the user interface from the overall
distribution of votes. Thus we can identify variation in users'
inclination to vote on a story they see.

We compared model predictions with observed number of votes, either
at the end of our sample for a story or two days after submission,
whichever was earlier.
For the promoted stories, the \rms\ relative error between the
number of votes and the model prediction is $14\%$, corresponding to
a \rms\ error of $109$ votes.
For stories not promoted these values are $14\%$ and $1.1$ votes,
respectively.

Our data set, examining the front page at one hour intervals, only
provides the hour within which the story was promoted. Based on this
coarse observation, our model tends to underestimate the promotion
time, with the median error of about 4 hours. This is likely due
mainly to the large variation in the number of fans for subsequent
voters: the model explicitly includes the submitter's fans but only
considers the average subsequent growth in fans with
Eq.~\ref{eq:Sv}.

\begin{figure}[t]
\begin{center}
\includegraphics[width=3in]{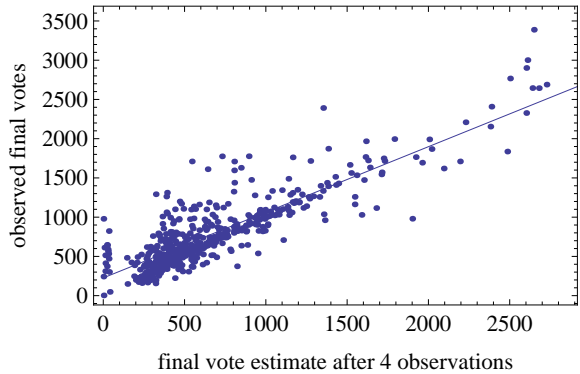}
\end{center}
\caption{Observed number of final votes for promoted stories
compared to prediction from the model using the first four
observations of each story to estimate the story's $r$ value. The
line is the best linear fit, with slope $0.84$.}\label{fig:early
estimates}
\end{figure}

The $r$ values can be  estimated from the early history of each
story. For instance, using just the first 4 observations for each
promoted story increases the relative error in the votes to $34\%$.
The predicted numbers of votes have $87\%$ correlation with the
observed numbers so early observations provide a strong prediction
of the relative ordering of numbers of votes stories will receive,
as illustrated in Fig.~\ref{fig:early estimates}.
This corresponds to the predictability of eventual ratings from
early reaction to new content seen on Digg and
YouTube~\cite{lerman08,szabo09}. Once a story reaches the front
page, its subsequent growth in votes is well-predicted from the
number of votes it receives shortly after promotion when accounting
for the hourly and daily variation in story submission
rate~\cite{szabo09}. However, predictions based on early
observations of the story prior to its promotion benefit from
accounting for the growth in fans~\cite{lerman07} as included in our
model. As a simple comparison, we determined the predicted number of
votes based on extrapolating from the rate a story accumulated votes
during the first 4 observations. This simpler model, which does not
consider the number of fans for the story's voters, has a lower
correlation, $75\%$, with the observed numbers and a larger \rms\
error. A randomization test comparing these two methods indicates
this reduction in performance is statistically significant
($p$-value less than $5\times 10^{-4}$). Thus our model,
incorporating the average growth in number of fans, provides a
better description of how stories accumulate votes than simply
extrapolating from early observations while on the upcoming pages.
More generally, by estimating the ``interestingness'' of a story
from early votes, we separate the influence of changing visibility
in the Digg user interface from the underlying rate at which users
will vote on the story if they see it.

The general behavior, of slow vote growth on the upcoming list,
mainly due to fans, followed by much faster growth if the story is
promoted, is qualitatively similar to previous stochastic models of
Digg~\cite{lerman07}. Thus these aggregate behaviors are somewhat
robust with respect to modeling details of individual user
behaviors. Our model improves on this prior work with a better
motivated derivation and incorporating a more principled description
of a story's visibility and the growth in the number of fans. We
also provide a more complete empirical analysis of how fans affect
story promotion.

\begin{figure}[t]
\begin{center}
\includegraphics[width=3in]{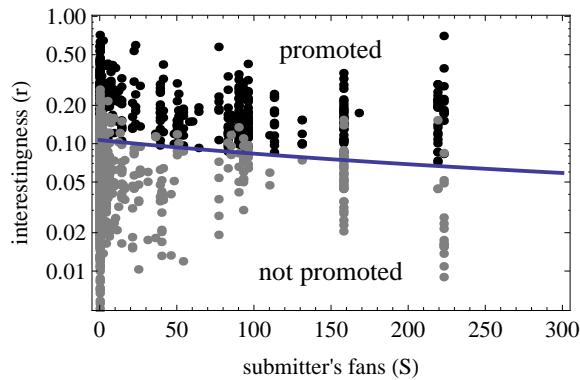}
\end{center}
\caption{Story promotion as a function of $S$ and $r$. The $r$
values are shown on a logarithmic scale. The model predicts stories
above the curve are promoted to the front page. The points show the
$S$ and $r$ values for the stories in our data set: black and gray
for stories promoted or not, respectively.}\label{fig:promotion}
\end{figure}

Fig.~\ref{fig:promotion} shows parameters required for a story to
reach the front page according to the model, and how that prediction
compares to the stories in our data set. The stories in
Fig.~\ref{fig:predictions} are all above the promotion threshold.
The model's prediction of whether a story is promoted is correct for
$95\%$ of the stories in our data set.
For promoted stories, the correlation between $S$ and $r$ is
$-0.13$, which is significantly different from zero ($p$-value less
than $10^{-4}$ by a randomization test).
Thus a story submitted by a poorly connected user (small $S$) tends
to need high interest (large $r$) to be promoted to the front
page~\cite{lerman07digg}.

Our model gives a reasonable qualitative account of how user
behavior leads to stories' promotion to the front page and the
eventual saturation in the number of votes they receive due to their
decreasing visibility. Additional properties of the interface and
user population could be added to the model for a more accurate
analysis of the behavior. For example, a submitter's fans may find
the story more interesting than the general Digg audience,
corresponding to different $r$ values for these groups of users.
Digg has additional mechanisms for presenting stories, e.g., by
topic and ``top stories'' that have received especially many votes
in recent time periods. Users finding stories via these parts of the
Digg interface could be included with additional terms in
Eq.~\ref{eq:diggs}.

We model users coming to Digg independently with unform rates $\nu$
and $\omega$. In fact, the rates vary systematically over hours and
days~\cite{szabo09}, and individual users have a wide range in time
between visits~\cite{vazquez06}. In our model, this variation gives
time-dependent values for $\nu$, describing the rate users come to
Digg, and $\frontPageGrowth$ and $\newPageGrowth$, which relate to
the rate new stories are posted and promoted. Another
time-dependence arises from the decreasing novelty of stories over
time~\cite{wu07}. In our model, this behavior would arise from a
combination of the story's reduced visibility and a time-dependent
decrease in the story's interest to users (i.e., a reduction in
$r$). Moreover, in some cases, the interestingness of web content,
not just its visibility, depends on how many votes it has
received~\cite{salganik06}. The model could include this behavior
with $r$ depending directly on $\voteTotal$. Additional dynamics
arises if users change behavior with experience on the web site, or
the site's algorithms or interface are modified. In this case, user
model parameters would need occasional reestimation.

These possibilities of incorporating additional details in the user
models illustrates how the stochastic approach provides insights
into how aggregate behavior arises from the users, in contrast to
models evaluating regularities only in the aggregate behaviors. In
particular, user models can help distinguish aggregate behaviors
arising from intrinsic properties of the stories (e.g., their
interestingness to the user population) from behavior due to the
information the web sites provides, such as ratings of other users
and how stories are placed in the site, i.e., visibility.

\section{Discussion}
\label{sec:discussion}

We described a general approach to relating simple models of user
choices to aggregate properties of systems involving many users.
Modeling user-participatory web sites is one application of this
approach, as we illustrated for Digg. Observations allow estimating
the rate parameters appearing in the model. Comparing solutions to
the model with observations can also help identify approaches to
improving the model, e.g., by including heterogeneous preferences
among users. The user state diagram is determined by the actions and
information the web site makes available to users. Whether this
approach results in a tractable model depends on the questions one
is interested in and how much user behavior depends on details of
user history or on the specific choices of other users rather than
just a few aggregate measures provided by the web site.

Although we focus on Digg, many user-contributory sites have similar
structural properties, including extended distributions of user
activity and graphical properties of their social
networks~\cite{newman03,wilkinson08}. Thus the stochastic formalism
relating user behavior to aggregate content rating should generalize
to other web sites.

The connection between user state transitions and aggregate behavior
allows investigation of how changes to the web site may change
aggregate behaviors. Such hypothetical uses of the modeling approach
can suggest improvements to the web site. For example, Digg's
promotion algorithm could take into account the number of fans a
submitter has, making it more difficult for highly connected users
to get uninteresting stories promoted to the front page.

This framework is particularly relevant when information on specific
users is limited, as is their set of actions (e.g., posting stories
and voting on them in Digg). The framework is less well suited to
describing complicated history-dependent actions (e.g., individual
users who remember how others treated them in the past as when
forming reputations in an e-commerce context). Moreover, while the
model can suggest how changes to underlying parameters or user
behaviors will affect overall observations, the model provides
correlations rather than causal connections between users and
observed behavior. In general, there could be other effects, not
included in the available observations of the users, that
significantly affect behavior and therefore may limit inference from
the model of changes that may achieve some more desired behavior
(e.g., users spending more time at a web site). Nevertheless, the
relations seen with stochastic models can suggest ways to improve
the behavior which could be tested, either directly through
experimental manipulation of the web site~\cite{salganik06} or
through smaller-scale experiments~\cite{kagel95}.

A practical challenge for using these models is identifying the
relevant states for the users and estimating the transition rates
among these states~\cite{brown03,ellner06}. To some extent, online
activities simplify this problem through their limited set of
actions and information provided to users. However, web sites can
become more personalized over time, e.g., with collaborative
filtering for recommendations based on history. This leads to more
history-dependence in user behavior and the open question of whether
the history-dependence can be summarized in simple additional state
variables for the user -- such as probability a recommendation is
relevant being a function of number of visits the person has had to
a site. If so, the model only requires a few additional state
variables -- in this case number of visits -- to regain the Markov
property. Alternatively, we can generalize the model to allow the
transition to the next state to depend not just on the current state
but also some fixed number of past states, as has been applied to
dynamic task allocation~\cite{lerman06ijrr}.

As web sites develop greater complexity and personalization,
model-based design tools could help identify aggregate consequences
of design choices of actions and information provided to users. More
broadly, such models could complement economic or game theory
analyses of the incentives for participation provided to the users.

\small
\section*{Acknowledgments} This work is based on research supported in part by the National Science Foundation under awards IIS-0535182 and
BCS-0527725. We thank M. Brzozowski and G. Szabo for helpful
comments.


\begin{thebibliography}{10}

\bibitem{brown03}
Kevin~S. Brown and James~P. Sethna.
\newblock Statistical mechanical approaches to models with many poorly known
  parameters.
\newblock {\em Physical Review E}, 68:012904, 2003.

\bibitem{clauset07}
Aaron Clauset, Cosma~Rohilla Shalizi, and M.~E.~J. Newman.
\newblock Power-law distributions in empirical data.
\newblock arxiv.org preprint 0706.1062, 2007.

\bibitem{courtois85}
P.~J. Courtois.
\newblock On time and space decomposition of complex structures.
\newblock {\em Communications of the ACM}, 28(6):590--603, June 1985.

\bibitem{ellner06}
Stephen~P. Ellner and John Guckenheimer.
\newblock {\em Dynamic Models in Biology}.
\newblock Princeton Univ. Press, Princeton, NJ, 2006.

\bibitem{galstyan05}
Aram Galstyan, Tad Hogg, and Kristina Lerman.
\newblock Modeling and mathematical analysis of swarms of microscopic robots.
\newblock In P.~Arabshahi and A.~Martinoli, editors, {\em Proc. of the IEEE
  Swarm Intelligence Symposium (SIS2005)}, pages 201--208, 2005.

\bibitem{hogg87PhysRep}
Tad Hogg and Bernardo~A. Huberman.
\newblock Artificial intelligence and large scale computation: A physics
  perspective.
\newblock {\em Physics Reports}, 156:227--310, 1987.

\bibitem{huberman98}
Bernardo~A. Huberman, Peter L.~T. Pirolli, James~E. Pitkow, and Rajan~M.
  Lukose.
\newblock Strong regularities in {World Wide Web} surfing.
\newblock {\em Science}, 280:95--97, 1998.

\bibitem{kagel95}
John Kagel and Alvin~E. Roth, editors.
\newblock {\em The Handbook of Experimental Economics}.
\newblock Princeton Univ. Press, 1995.

\bibitem{vankampen92}
N.~G.~Van Kampen.
\newblock {\em Stochastic Processes in Physics and Chemistry}.
\newblock Elsevier Science, Amsterdam, revised and enlarged edition, 1992.

\bibitem{lerman06ijrr}
K.~Lerman, Chris~V. Jones, A.~Galstyan, and Maja~J. Matari{\'c}.
\newblock Analysis of dynamic task allocation in multi-robot systems.
\newblock {\em International Journal of Robotics Research}, 25(3):225--242,
  2006.

\bibitem{lerman07}
Kristina Lerman.
\newblock Social information processing in social news aggregation.
\newblock {\em IEEE Internet Computing: special issue on Social Search},
  11(6):16--28, 2007.

\bibitem{lerman07digg}
Kristina Lerman.
\newblock Social networks and social information filtering on {Digg}.
\newblock In {\em Proc. of International Conference on Weblogs and Social Media
  (ICWSM-07)}, 2007.

\bibitem{lerman08}
Kristina Lerman and Aram Galstyan.
\newblock Analysis of social voting patterns on {Digg}.
\newblock In {\em Proceedings of the 1st ACM SIGCOMM Workshop on Online Social
  Networks}, pages 7--12, New York, 2008. ACM.

\bibitem{lerman01}
Kristina Lerman, Aram Galstyan, Alcherio Martinoli, and Auke~Jan Ijspeert.
\newblock A macroscopic analytical model of collaboration in distributed
  robotic systems.
\newblock {\em Artificial Life}, 7:375--393, 2001.

\bibitem{martinoli04}
A.~Martinoli, K.~Easton, and W.~Agassounon.
\newblock Modeling of swarm robotic systems: A case study in collaborative
  distributed manipulation.
\newblock {\em Int. Journal of Robotics Research}, 23(4):415--436, 2004.

\bibitem{newman03}
M.~E.~J. Newman.
\newblock The structure and function of complex networks.
\newblock {\em SIAM Review}, 45(2):167--256, 2003.

\bibitem{opper01}
Manfred Opper and David Saad, editors.
\newblock {\em Advanced Mean Field Methods: Theory and Practice}.
\newblock MIT Press, Cambridge, MA, 2001.

\bibitem{robins07}
Garry Robins, Pip Pattison, Yuval Kalish, and Dean Lusher.
\newblock An introduction to exponential random graph (p*) models for social
  networks.
\newblock {\em Social Networks}, 29:173--191, 2007.

\bibitem{salganik06}
Matthew~J. Salganik, Peter~Sheridan Dodds, and Duncan~J. Watts.
\newblock Experimental study of inequality and unpredictability in an
  artificial cultural market.
\newblock {\em Science}, 311:854--856, 2006.

\bibitem{simon96}
Herbert~A. Simon.
\newblock {\em The Sciences of the Artificial}.
\newblock MIT Press, Cambridge, MA, 3rd edition, 1996.

\bibitem{steglich07}
Christian Steglich, Tom A.~B. Snijders, and Michael Pearson.
\newblock Dynamics networks and behavior: Separating selection from influence.
\newblock Technical report, Interuniversity Center for Social Science Theory
  and Methodology, July 2007.

\bibitem{szabo09}
Gabor Szabo and Bernardo~A. Huberman.
\newblock Predicting the popularity of online content.
\newblock Technical report, HP Labs, Nov. 2008.
\newblock Available at hpl.hp.com/research/scl/papers/predictions.

\bibitem{vazquez06}
A.~V\'azquez, J.~Gama Oliveira, Z.~Dezso, K.-I. Goh, I.~Kondor, and A.-L.
  Barabasi.
\newblock Modeling bursts and heavy tails in human dynamics.
\newblock {\em Physical Review E}, 73:036127, 2006.

\bibitem{wilkinson08}
Dennis~M. Wilkinson.
\newblock Strong regularities in online peer production.
\newblock In {\em Proc. of the 2008 ACM Conference on E-Commerce}, pages
  302--309, 2008.

\bibitem{wu07}
Fang Wu and Bernardo~A. Huberman.
\newblock Novelty and collective attention.
\newblock {\em Proc. of the Natl. Acad. Sci.}, 104:17599--17601, 2007.

\end{thebibliography}
\end{document}